\documentclass[aps,pra,preprint]{revtex4-2}
\usepackage{graphicx}
\usepackage{color}
\usepackage{amsmath,amssymb}
\usepackage{bigints}
\usepackage{comment}
\usepackage{hyperref}
\usepackage{bm}
\begin{document}

\title{Mpemba as an Emergent Effect of System Relaxation}

\author{Gourab Das}
\email{gourabdasofficial@gmail.com}

\affiliation{Department of Physical Sciences, Indian Institute of Science Education and Research Kolkata, Mohanpur 741246, India}

\begin{abstract}
\noindent The Mpemba effect (MpE), where a far-from-equilibrium state of a system relaxes faster compared to a state closer to it, is a well-known counterintuitive phenomenon in classical and quantum systems. Various system-specific theories have been proposed to explain this anomalous behavior in driven systems, though the fundamental mechanism of MpE in undriven systems, where MpE was first observed, remains unresolved. This paper provides a generic model of MpE for a quantum spin-$1/2$ system following Markovian relaxation dynamics, regardless of system structure or environment. The key lies in the overlap of initial states with the fast relaxation mode; here, the constituents create a fast decay mode via interaction through the shared environment to show MpE, indicating that MpE happens due to the collective behavior of the system. I also show that a system with anisotropic relaxation naturally exhibits MpE, even without a shared environment among the particles.
\end{abstract}


\maketitle

\section*{Introduction}

Erasto B. Mpemba rediscovered the counterintuitive phenomenon where a hot liquid mixture freezes earlier than an identical cold liquid mixture \cite{Mpemba_1969}, though this intriguing phenomenon had been mentioned earlier in various texts throughout history \cite{Aristotle_1962, Bacon_1902, Descartes_1956, Groves_1962, Bacon_1962}. The Mpemba effect (MpE), as coined by the community, describes an umbrella of phenomena where a far-from-equilibrium state equilibrates faster than a state that is closer to the equilibrium state. The works of Mpemba and Osborne \cite{Mpemba_1969} and Kell \cite{Kell_1969} sparked huge interest in studying MpEs in various classical and quantum systems \cite{Chaddah_2010, Brownridge_2011, Greaney_2011, Ahn_2016, Lasanta_2017, Keller_2018, Hu_2018, Baity_2019, Baity_2019, Biswas_2020, Kumar_2020, Busiello_2021, Biswas_2023, Teza_2025, Ares_2025}. Recently, MpE has been observed experimentally in quantum systems \cite{Zhang_2025, Aharony_2024}. MpEs are known to be useful for quantum batteries \cite{Medina_2025}, refrigerators \cite{Mondal_2025}, etc.

\begin{figure}
\includegraphics[width=0.7\linewidth]{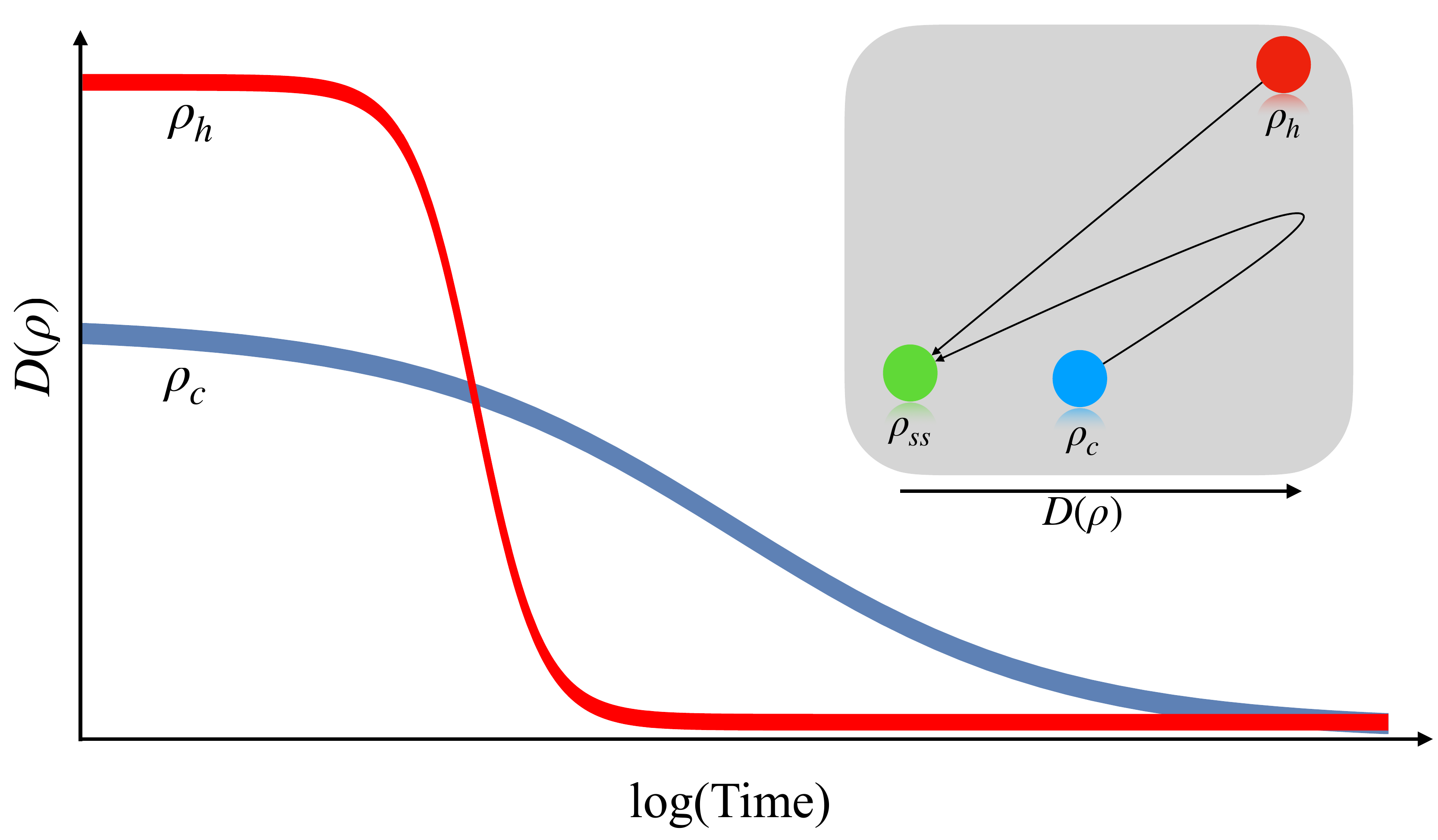}
\caption{(color online)
A schematic of MpE, where a far-from-equilibrium state ($\rho_h$) relaxes faster than a state closer ($\rho_c$) to equilibrium ($\rho_{ss}$). Here, $D$ represents a metric to measure distance; crossing of the trajectories indicates MpE in action.
}
\label{fig:1}
\end{figure}

However, the existence of MpE is also being questioned in various conditions \cite{Vynnycky_2010, Burridge_2016, Burridge_2020, Bechhoefer_2021}, despite being studied over more than half a century \cite{Gal_2020, Schwarzendahl_2022, Holtzman_2022, Teza_2023, Pem_2024, Santos_2024, Van_2025, Teza_2025}. Another major challenge MpE faces is the lack of a universally accepted explanation; various theories exist in different contexts to explain this anomalous behavior. In classical systems, MpE is often attributed to memory effects, although Markovian classical systems are known to exhibit MpE \cite{Lu_2017}. Various mechanisms have been proposed to explain MpE in quantum systems too, e. g. bath engineering \cite{Chatterjee_2023, Wang_2024, Moroder_2024, Strachan_2025}, the quasi-particles picture \cite{Ares_2023}, unitary rotation to initial states \cite{Carollo_2021, Kochsiek_2022, Bao_2025}, potential energy surface explanations \cite{Lu_2017, Nava_2024, Liu_2025}, etc. These studies focus on the process of creating a fast decay mode following the mentioned mechanisms; the far-from-equilibrium state following the fast decay mode relaxes quickly, while the closer state, following the slow decay mode, takes more time to equilibrate, showing MpE (Fig. \ref{fig:1}). These articles suggest that there is a broad umbrella of system-specific mechanisms that show this anomalous relaxation. Though these studies are able to explain MpE for driven quantum systems, MpE was originally shown for undriven systems during relaxation, the cause of which remains unresolved\cite{Mpemba_1969}. 

The origin of MpE in Markovian systems remains poorly understood. Although experimental realizations of quantum MpEs have been achieved in Markovian settings, where systems relax through engineered decay channels \cite{Chalas_2024, Aharony_2024}, a general mechanism underlying MpE during relaxation dynamics is still lacking. Establishing such a mechanism is important because it could provide a unified framework applicable to both open quantum and classical systems. In this work, I present a generic model of an undriven spin-$1/2$ open quantum system that exhibits MpE under Markovian relaxation dynamics. The model assumes all-to-all coupling between constituents, independent of their spatial positions, with the interaction taken to be a generic energy-conserving coupling. Within a mean-field treatment, I show that the coupling strength does not influence the emergence of MpE, implying that the analysis also applies to systems of noninteracting particles. This insensitivity originates from the azimuthal symmetry of the relaxation dynamics. I further demonstrate that MpE emerges when the system constituents interact with a common environment. In this case, the shared environment mediates collective interactions that generate a fast relaxation mode in addition to slower decay processes. This finding suggests that collective relaxation is the physical mechanism responsible for MpE in otherwise isotropic systems. While previous studies typically assumed the existence of distinct fast and slow relaxation modes, this work identifies their microscopic origin. Specifically, collective dissipation induced by a shared environment naturally produces a fast relaxation mode whose decay rate is enhanced by the particle number, thereby creating and amplifying a separation of relaxation timescales. I also investigate systems with anisotropic relaxation. In contrast to isotropic systems, anisotropic relaxation intrinsically generates multiple relaxation timescales, leading to the coexistence of fast and slow decay modes. As a result, such systems can exhibit MpE even in the absence of collective relaxation mediated by a shared environment.

\section*{System}

\begin{figure}[h!]
\hspace*{-2mm}\raisebox{4.6cm}{(a)}\hspace*{-1.4mm}
\includegraphics[width=0.47\linewidth]{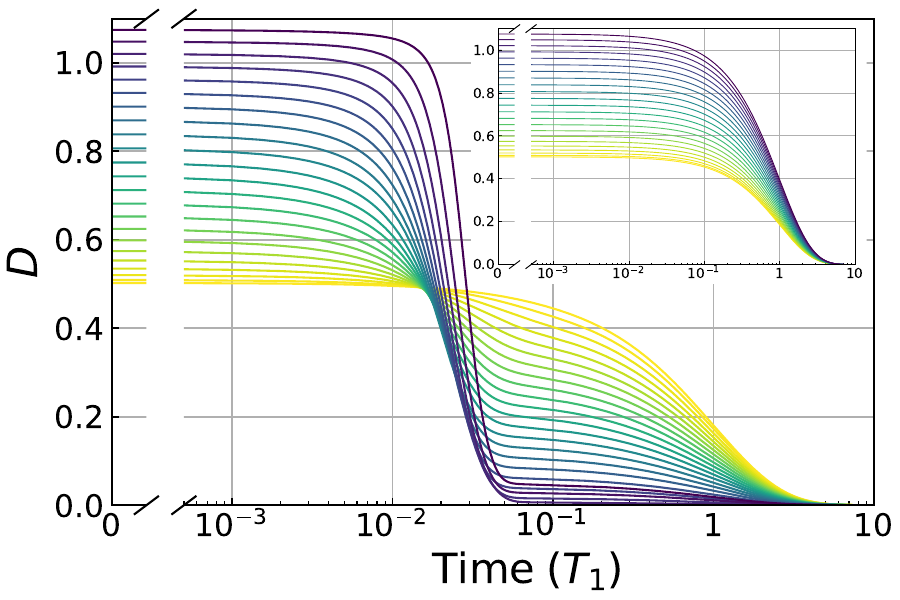}
\hspace*{-2mm}\raisebox{4.6cm}{(b)}\hspace*{-1.4mm}
\includegraphics[width=0.47\linewidth]{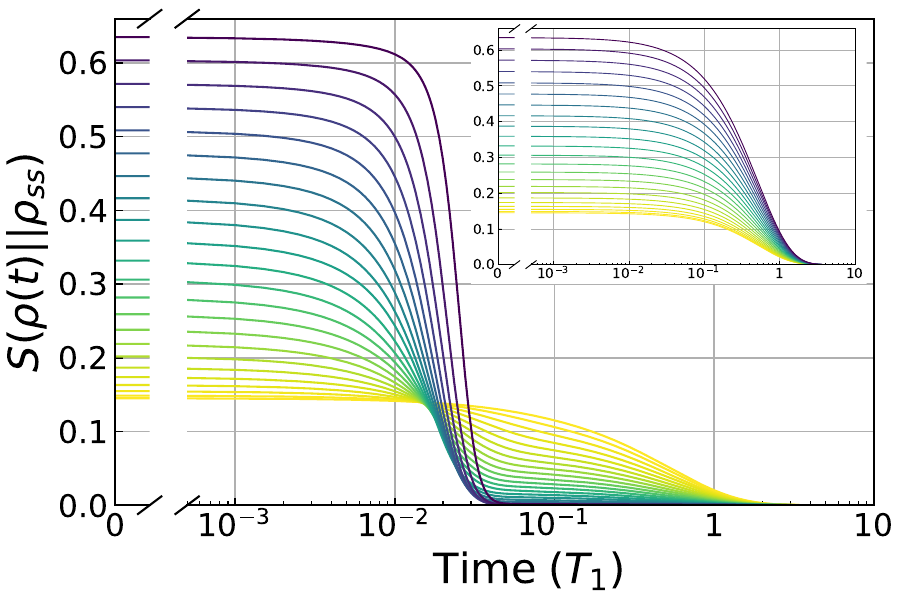}
\caption{(color online)
MpE in action. The evolution of metrics (a) $D$ and (b) $S(\rho (t) || \rho_{ss})$, with respect to equilibrium, for different initial states; these show crossover trajectories, \textit{i.e.}, MpE.
The colors represent the initial state's distance (or relative entropy) from the equilibrium state. The insets show the evolution of the same for a system without a shared environment; in this case, we do not see MpE.
The parameters used here are $T_2 = T_1$, $n = 10^3$, and $M_\circ = 0.5$; while $r$ and $\theta$ are chosen from the line $-\frac{5}{6} r + \frac{\theta}{\pi} = \frac{1}{2}$. The states on this line possess different overlaps with the slow ($r$) and fast ($\theta$) eigenmodes. Since the relative contributions of these two modes vary from state to state, their relaxation rates are not identical. This variation in relaxation rates among the states on the line gives rise to the observed behavior here.
}
\label{fig:2}
\end{figure}

Let's consider our system consisting of an ensemble of $n$ spin-half particles coupled among themselves via the following Hamiltonian,
\begin{equation}
\mathcal{H} = J_\circ \sum_{i<j} \sigma_i^z \sigma_j^z + \left( J_1 \sum_{i=1}^n \sum_{r=1}^{n-1} \sigma_i^+ \sigma_{i+r}^- + h.c. \right) \quad \text{with} \;\; i+r \equiv \left( \left(i+r-1\right) \text{mod}\; n \right) + 1,
\label{eq:1}
\end{equation}
where $h.c.$ represents the Hermitian conjugate of the previous term and $\sigma_i$ being the Pauli matrices on the $i$-th particle of the ensemble. Note, this coupling Hamiltonian conserves excitations (\textit{i.e.} it is energy-conserving Hamiltonian) and incorporates Ising, XX, XXZ, XY, isotropic Heisenberg, secular dipolar, Dzyaloshinskii–Moriya couplings etc. Here, the particles are considered to be all-to-all coupled with equal strength irrespective of their position for the simplicity of the model. The system is connected to a generic bosonic environment such that the constituents share the same environment. Hence, the density matrix of the system relaxes following the Lindblad master equation, in the natural unit \cite{Breuer_2002},
\begin{eqnarray}
\frac{d \rho}{dt} &=& -i \left[ \mathcal{H} , \rho \right] + \sum_{i=1}^n \sum_{j=1}^n \mathcal{D}_{ij} \left[ \rho \right], \nonumber \\
\mathcal{D}_{ij} \left[ \rho \right] &\equiv& \sum_{k=1}^3 \left( L_k(i) \rho L_k(j)^\dagger - \frac{1}{2} \left\{ L_k(j)^\dagger L_k(i), \rho \right\} \right)
\label{eq:2}
\end{eqnarray}
where, $L_1(i) = \sqrt{\frac{1+M_\circ}{2T_1}} \sigma^+_i $, $L_2(i) = \sqrt{\frac{1-M_\circ}{2T_1}} \sigma^-_i $, $L_3(i) = \sqrt{\frac{1}{2T_\phi}} \sigma^z_i $ and $M_\circ$ is the environment's magnetization along $z$. Here, $T_1$ and $T_{\phi}$ represent the relaxation and dephasing timescales, respectively. Therefore, under mean-field approximations, the average magnetization obeys the Bloch equations as follows (in polar coordinates),
\begin{eqnarray}
\frac{dr}{dt} &=& \frac{M_\circ \cos \theta}{T_1} - r \left( \frac{\cos^2 \theta}{T_1} + \frac{\sin^2 \theta}{T_2} \right) \nonumber \\
\frac{d\theta}{dt} &=& -\sin\theta \left( \frac{n-1}{2T_1} M_\circ r + \cos \theta \left( \frac{1}{T_2} - \frac{1}{T_1} \right) + \frac{M_\circ}{r T_1} \right) \nonumber \\
\frac{d\phi}{dt} &=& (n-1) \left( 2 J_\circ - Re[J_1]\right) r \cos \theta\label{eq:3}
\end{eqnarray}
where, $T_2$ is the decoherence timescale of the system with $T_2^{-1} = 0.5 T_1^{-1} + T_\phi^{-1}$; here $T_1$ and $T_2$ have their standard representations \cite{Breuer_2002}. Note, the first term in the $\theta$ equation incorporates the decoherence effect due to the shared environment, and the equation for $\phi$ takes into account of the couplings among the particles. Here, the magnetization relaxes to $(r, \theta) = (M_\circ, 0)$, which corresponds to the equilibrium density matrix, $\rho_{ss}$; note that at this point $\phi$ is ill-defined. Therefore, the relaxation is supposed to be isotropic if $T_1 = T_2$; we will see in the next section that MpE emerges naturally for systems with anisotropic relaxation. Let's discuss how a system with as isotropic relaxation shows MpE; I use the Euclidean distance, $D = \sqrt{\left( r(t) \cos \theta (t) - M_\circ \right)^2 + r(t)^2 \sin^2 \theta(t)}$, of the magnetization and the relative entropy, $S(\rho(t) || \rho_{ss}) = \text{Tr} \left( \rho(t) \ln \left( \rho(t) / \rho_{ss} \right)\right)$, with respect to $\rho_{ss}$, to observe MpE, Fig. \ref{fig:2}. In this case, $D$ is proportional to the trace distance and Frobenius distance \cite{Ares_2025}. Fig. \ref{fig:2} shows crossover trajectories when particles of the system share the same environment, indicating the origin of MpE as the collective behavior of the constituents. Note, $D$ and $S$ are invariant under rotation about the $z$-axis due to the azimuthal symmetry of the relaxation processes. Therefore, the evolution in $\phi$ is ignored in the following discussions, unless explicitly mentioned, and the analysis can be generalized to systems independent of the particles' coupling strength (including the class of systems made of noninteracting particles as well), as the $r$ and $\theta$ equations of Eq. \eqref{eq:3} do not depend on the coupling strength.

\begin{figure}[h!]
\hspace*{-2mm}\raisebox{3.2cm}{(a)}\hspace*{-1.4mm}
\includegraphics[width=0.48\linewidth]{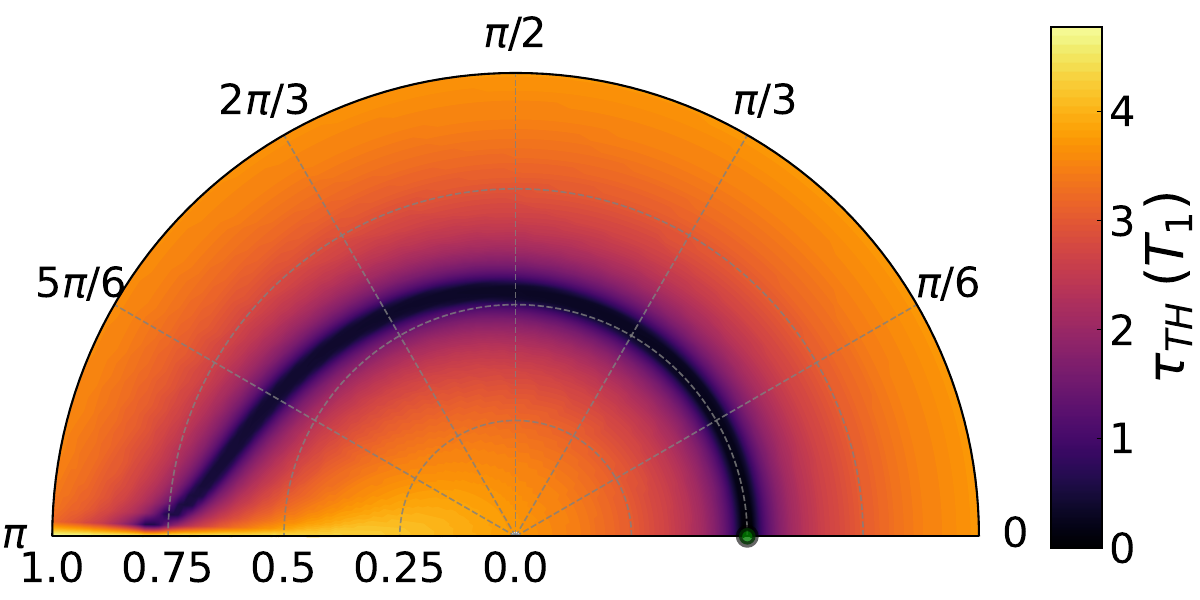}
\hspace*{-2mm}\raisebox{3.2cm}{(b)}\hspace*{-1.4mm}
\includegraphics[width=0.455\linewidth]{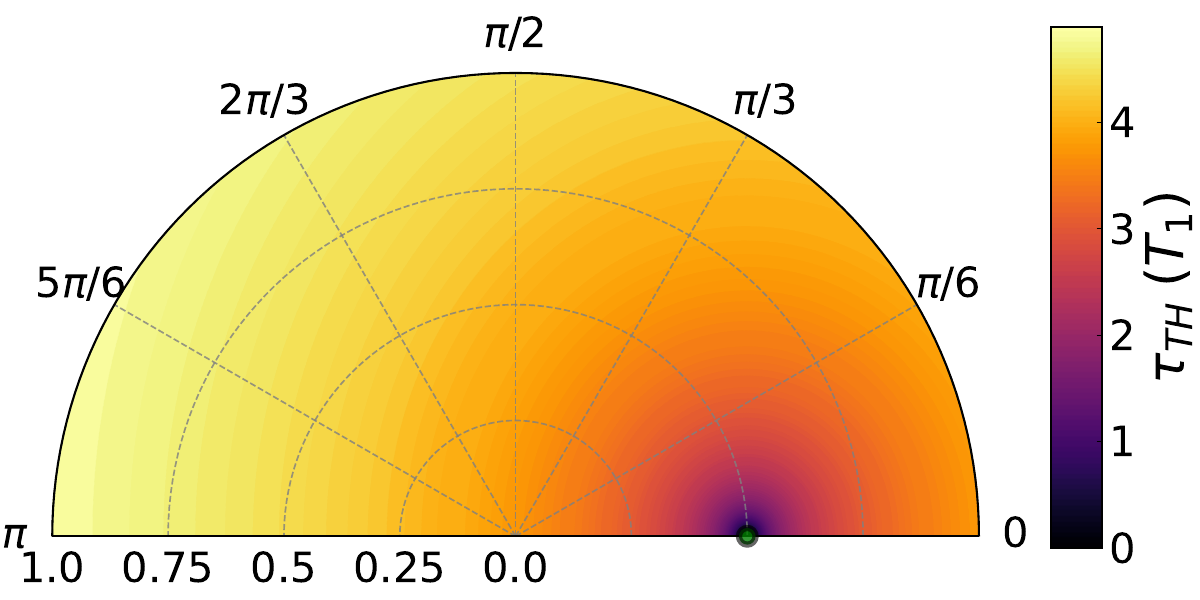}

\hspace*{-2mm}\raisebox{3.2cm}{(c)}\hspace*{-1.4mm}
\includegraphics[width=0.47\linewidth]{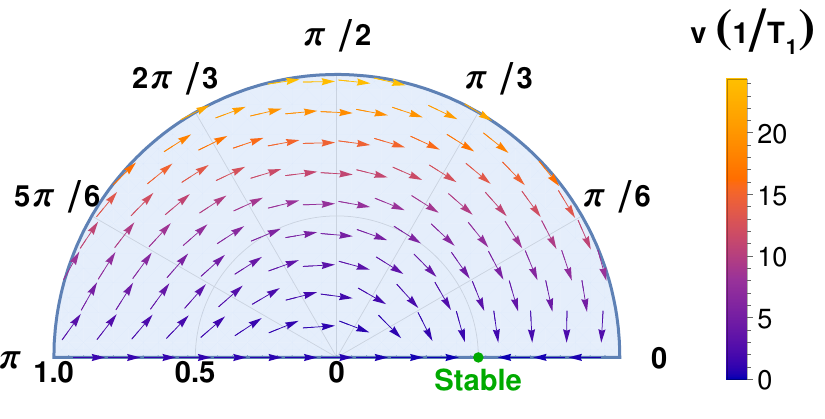}
\hspace*{-2mm}\raisebox{3.2cm}{(d)}\hspace*{-1.4mm}
\includegraphics[width=0.47\linewidth]{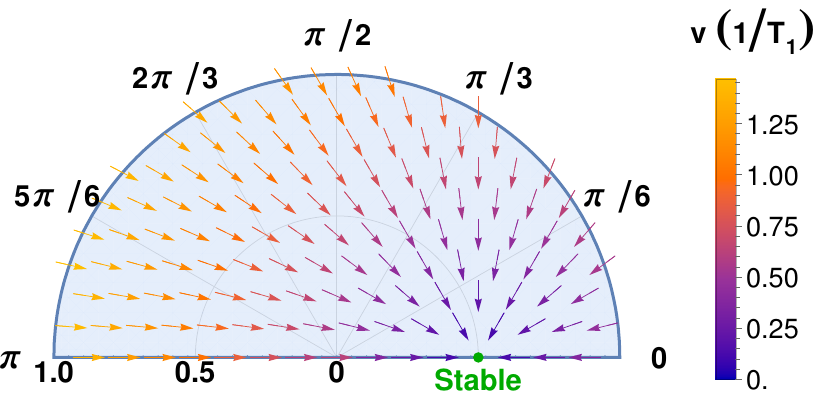}
\caption{(color online)
(a) and (b) show thermalization time, $\tau_{TH}$, for different initial states for systems with and without a shared environment, respectively (in polar coordinates). $\tau_{TH}$ is numerically calculated as when $S(\rho(t) || \rho_{ss})$ becomes less than a cutoff ($10^{-4}$ here). Equivalently, one may use the Euclidean distance $D$ instead of the relative entropy $S$ to determine the thermalization time $\tau_{TH}$. This would yield qualitatively similar graphs, although not identical, owing to the different functional forms of $D$ and $S$. The equivalence between $D$ and $S$ for spin-$1/2$ systems is demonstrated in Appendix~\ref{A:1}. Here, the green dot represents the equilibrium state, $\rho_{ss}$. The states closer to $\rho_{ss}$ thermalize earlier in (b), \textit{i.e.}, no MpE; while states with completely different initial states thermalize in (a), thus showing MpE. 
(c) and (d) represent the trajectories of states to $\rho_{ss}$ for systems with and without a shared environment, respectively. $\rho_{ss}$ is the stable fixed point for the system, represented here by the green dot. The color of the arrows denotes the speed of relaxation, $v$, following the trajectories. The speed increases with the distance from $\rho_{ss}$ and the trajectories are radial towards $\rho_{ss}$ in (d); whilst in (c), the trajectories are curved and speed increases with the distance from $\theta = 0$ or $\theta = \pi$ lines. Thus, (c) and (d) explain the behaviors seen in (a) and (b), respectively.
Here, the used parameters are  $T_2 = T_1$, $n = 10^2$, and $M_\circ = 0.5$.
}
\label{fig:3}
\end{figure}

Eq. \eqref{eq:3}, ignoring the equation for $\phi$, has only one fixed point at $(r, \theta) = (M_\circ, 0)$ and it is stable, as the corresponding Jacobian takes the following form,
\begin{equation}
\setlength{\arraycolsep}{10pt}
J = \begin{bmatrix}
-\frac{\cos^2 \theta}{T_1}-\frac{\sin^2 \theta}{T_2} & -\frac{M_\circ \sin \theta}{T_1}+r\sin 2\theta \left( \frac{1}{T_1} - \frac{1}{T_2} \right) \\
\left(\frac{1}{r^2} - \frac{n-1}{2}\right) \frac{M_\circ}{T_1} \sin \theta & -\frac{M_\circ}{T_1}\cos\theta \left(\frac{1}{r} + \frac{n-1}{2} r\right)+\cos 2\theta  \left( \frac{1}{T_1} - \frac{1}{T_2} \right)
\end{bmatrix}
\label{eq:4}
\end{equation}
with eigenvalues $\lambda_1 = -T_1^{-1}$ and $\lambda_2 = -\left(T_2^{-1} + 0.5(n-1)M_\circ^2T_1^{-1}\right)$ at the fixed point, corresponding to the $r$ and $\theta$ directions, respectively \cite{Strogatz_2024}. Since the dynamics considered here is continuous, the Lyapunov exponents coincide with the Jacobian eigenvalues at the fixed point \cite{Strogatz_2024}. Consequently, $|\lambda_{1(2)}|$ determines the relaxation rate along the $r$ ($\theta$) direction. The two eigendirections therefore relax at different speeds, as illustrated in Fig. \ref{fig:3}(c). This separation of relaxation rates provides the basis for MpE. States with negligible overlap with the slower mode relax predominantly through the faster mode and can reach equilibrium earlier even when they start farther from it. As a result, trajectories of $D$ or $S(\rho(t)\Vert\rho_{ss})$ exhibit a crossover, with the characteristic MpE timescale set by the minimum of the $|\lambda_{1(2)}|^{-1}$. Since typically $T_1 \geq T_2$, the faster mode is associated with $\lambda_2$, and thus $|\lambda_2|^{-1}$ determines the MpE timescale (Fig. \ref{fig:2}). Importantly, this timescale decreases with increasing particle number $n$. As $n$ increases, trajectories become increasingly aligned with the $\theta$ direction, and states with $r=M_\circ$ form the first set of states to relax, as shown in Fig. \ref{fig:3}(a). This observation is consistent with the fact that MpE does not occur for all initial states \cite{Kumar_2020}. Because relaxation along the $\theta$ direction is faster than along the $r$ direction, states lying on the contour centered at the origin and passing through the equilibrium point $(r,\theta)=(M_\circ,0)$ have minimal overlap with the slow $r$ mode. These states therefore relax most rapidly and exhibit MpE.

The distance of an initial state from equilibrium can be visualized by the semicircular contours centered at the equilibrium point, whose radii directly represent the initial distance from equilibrium. A state with a larger projection onto the fast $\theta$ mode can therefore relax more quickly even when it starts on a contour of larger radius than another state that is initially closer to equilibrium but retains a finite overlap with the slow $r$ mode, as illustrated in Fig. \ref{fig:3}(a,b). The shared environment enhances the relaxation rate along the $\theta$ direction, generating a fast collective relaxation mode whose speed increases with $n$, thereby giving rise to MpE. Hence, MpE becomes increasingly pronounced as the particle number grows.

For systems where the constituents do not share the same environment, $\lambda_{1(2)} = -T_{1(2)}^{-1}$, implying an isotropic speed of convergence (for $T_1 = T_2$) that increases with the distance from the equilibrium state, and the trajectories are radially convergent towards equilibrium, Fig. \ref{fig:3}(d). Therefore, as a far-initial-state comes close to equilibrium following the trajectories, its relaxation speed decreases, eventually matching that of an initial state closer to equilibrium. Hence, the states close to equilibrium relax earlier, and we do not see MpE, as seen in the insets of Fig. \ref{fig:2}(a, b) and Fig. \ref{fig:3}(b). 

\section*{Discussions}

\begin{figure}[h!]
\hspace*{-2mm}\raisebox{4.6cm}{(a)}\hspace*{-1.4mm}
\includegraphics[width=0.47\linewidth]{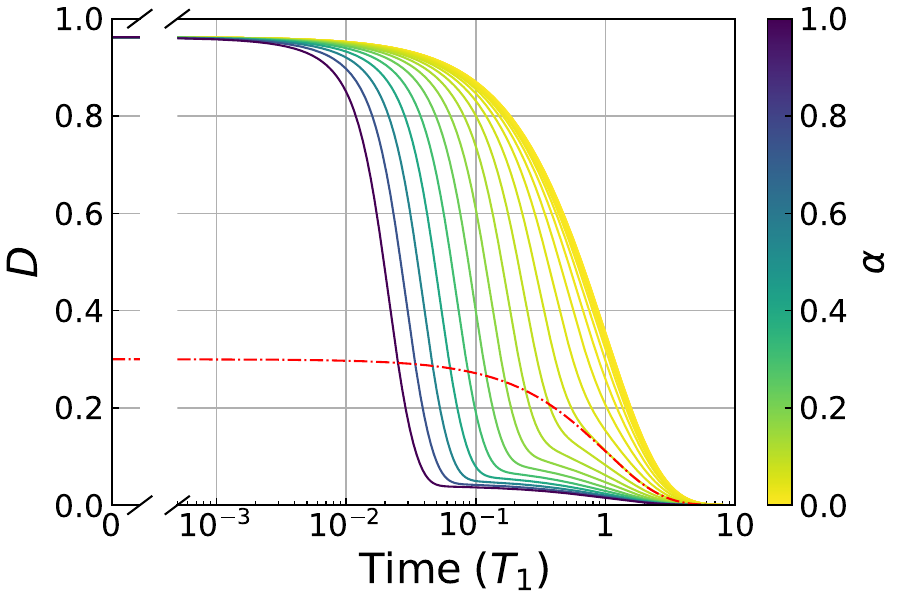}
\hspace*{-2mm}\raisebox{4.6cm}{(b)}\hspace*{-1.4mm}
\includegraphics[width=0.47\linewidth]{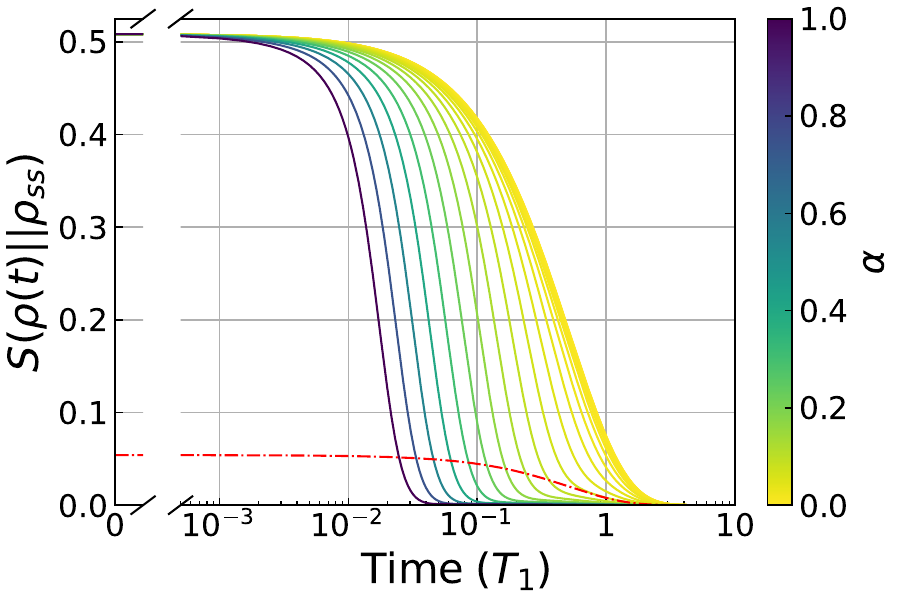}
\caption{(color online)
The effect of shared environments (among the particles) on MpE. The evolution of metrics (a) $D$ and (b) $S(\rho (t) || \rho_{ss})$, with respect to equilibrium, for different sharedness, $\alpha$. The solid lines are for the initial state $(r, \theta) = (\frac{19}{40}, \frac{43}{48}\pi)$, a initially further state, while the red dash-dot line is for a initially closer state with $(r, \theta) = (0.2, 0)$, whose trajectory remains unaffected for the changes in $\alpha$. These graphs show that the system become more prone to show MpE with increasing $\alpha$, \textit{i.e.}, the collective dissipation of the particles causes MpE.
The parameters used here are $T_2 = T_1$, $n = 10^3$, and $M_\circ = 0.5$.
}
\label{fig:4}
\end{figure}

Let us consider that there exists a parameter $\alpha$ that captures how the particles share the environment; $\alpha$ ranges from $0$ to $1$, with $\alpha = 0 $ meaning the particles have identical separate environments, and $\alpha = 1 $ representing the case where all the particles share the same environment. Therefore, the density matrix follows the following Lindblad master equation,
\begin{eqnarray}
\frac{d \rho}{dt} = -i \left[ \mathcal{H} , \rho \right] + \sum_{i=1}^n \left( \mathcal{D}_{ii} \left[ \rho \right] + \alpha \sum_{j=1, j\neq i}^n \mathcal{D}_{ij} \left[ \rho \right] \right)
\label{eq:5}
\end{eqnarray}
Fig. \ref{fig:4} shows that with increasing sharedness, $\alpha$, the system becomes more prone to show MpE. Hence, it verifies that the collective dissipation of the system constituents causes the MpE.

\begin{figure}[h!]
\hspace*{-2mm}\raisebox{4.6cm}{(a)}\hspace*{-1.4mm}
\includegraphics[width=0.47\linewidth]{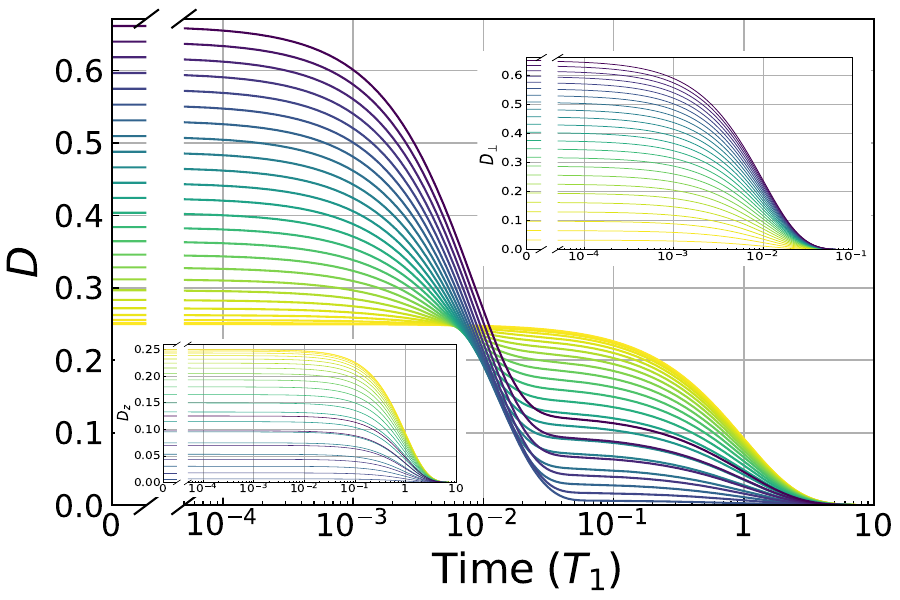}
\hspace*{-2mm}\raisebox{4.6cm}{(b)}\hspace*{-1.4mm}
\includegraphics[width=0.47\linewidth]{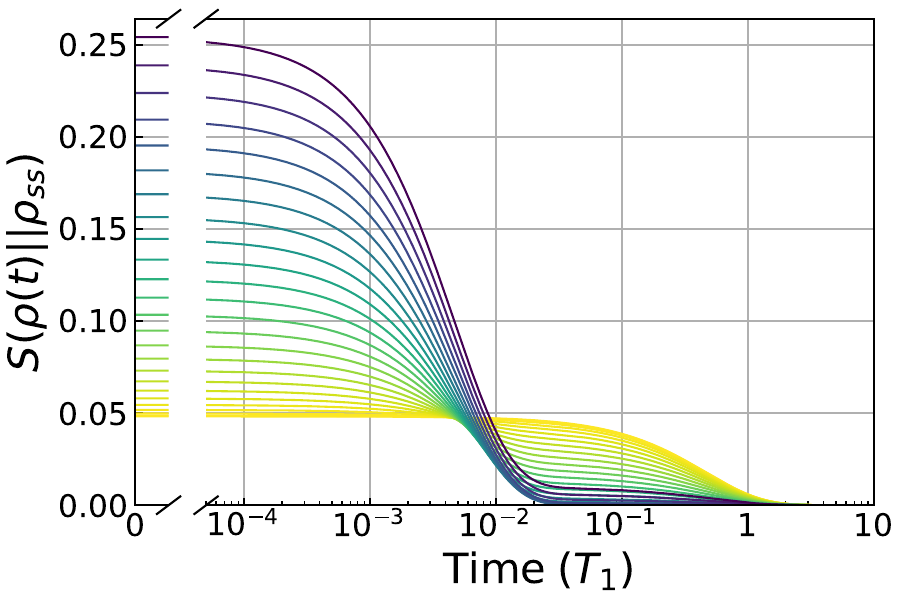}

\hspace*{-2mm}\raisebox{3.2cm}{(c)}\hspace*{-1.4mm}
\includegraphics[width=0.47\linewidth]{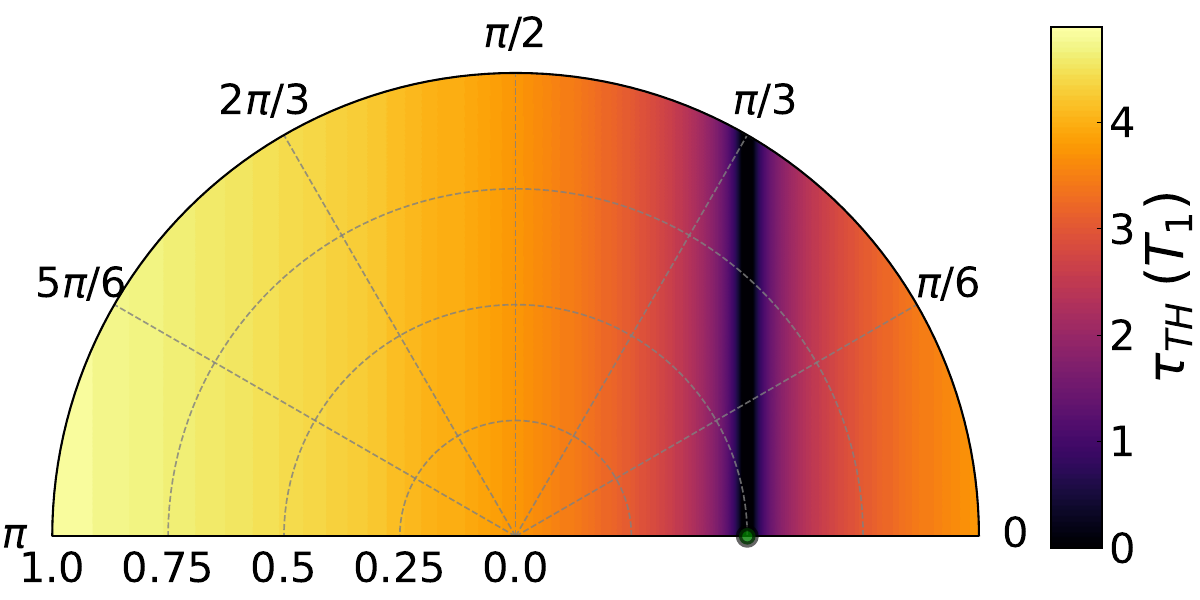}
\hspace*{-2mm}\raisebox{3.2cm}{(d)}\hspace*{-1.4mm}
\includegraphics[width=0.47\linewidth]{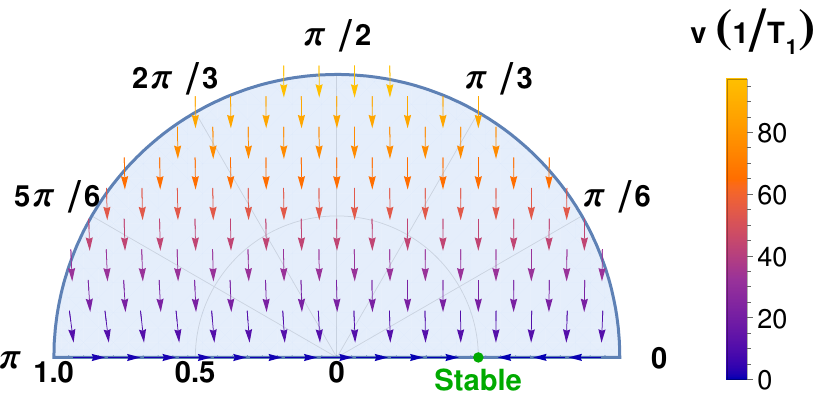}
\caption{(color online)
MpE seen using both the metrics (a) $D$ and (b) $S(\rho(t) || \rho_{ss})$. Though $D_z$ or $D_\perp$ do not show crossover trajectories. The colors represent the initial state’s distance (or relative entropy) from $\rho_{ss}$; $r = 0.75$ and $\theta$ is varied between $0$ and $\pi/3$ to generate (a) and (b). 
(c) shows that $\tau_{TH}$ does not depend on the initial state's distance from $\rho_{ss}$ (green dot), rather on the distance from the $r \cos \theta = M_\circ$ line. The trajectories of states, with colors representing the speed following them, are shown in (d), which explains the behavior of (c).
The used parameters are $T_2 = 0.01 T_1$ and $M_\circ = 0.5$.
}
\label{fig:5}
\end{figure}

Consider another system that consists of $n$ spin-halves coupled via the same coupling Hamiltonian as in Eq. \eqref{eq:1}, where the constituents are in contact with different local environments; for simplicity, let's take all these environments as identical, \textit{i.e.}, they have the same $M_\circ$ and the particles have the same relaxation timescales $T_1$ and $T_2$. Hence, the system's relaxation can be described by the following Lindblad master equation,
\begin{equation}
\frac{d \rho}{dt} = -i \left[ \mathcal{H} , \rho \right] + \sum_{i=1}^n \mathcal{D}_{ii} \left[ \rho \right]
\label{eq:6}
\end{equation}
where the dissipators $\mathcal{D}_{ij}$ are of the same form as mentioned earlier. Using the mean-field approximations, one gets the following set of equations for the average magnetization as earlier (in polar coordinates),
\begin{eqnarray}
\frac{dr}{dt} &=& \frac{M_\circ \cos \theta}{T_1} - r \left( \frac{\cos^2 \theta}{T_1} + \frac{\sin^2 \theta}{T_2} \right) \nonumber \\
\frac{d\theta}{dt} &=& -\sin\theta \left( \cos \theta \left( \frac{1}{T_2} - \frac{1}{T_1} \right) + \frac{M_\circ}{r T_1} \right) \nonumber \\
\frac{d\phi}{dt} &=& (n-1) \left( 2 J_\circ - Re[J_1]\right) r \cos \theta\label{eq:7}
\end{eqnarray}
Similar to the earlier case, we can ignore the variation in $\phi$ due to the azimuthal symmetry of the relaxation process. Thus, the next analysis becomes independent of coupling strength also. Hence, ignoring the evolution in $\phi$, the analytical solutions can be found as,
\begin{eqnarray}
r(t) = \sqrt{ r_z^2 + r_\perp^2} ; \quad \theta(t) = \tan^{-1} \left( \frac{r_\perp}{r_z} \right) \label{eq:8}
\end{eqnarray}
where, $r_z = M_\circ \left( 1-e^{-t/T_1}\right) + r_\circ \cos \theta_\circ e^{-t/T_1}$ and $r_\perp = r_\circ \sin \theta_\circ e^{-t/T_2}$ with $(r(0), \theta(0)) = (r_\circ, \theta_\circ)$. Note, Eq. \eqref{eq:8} is independent of particle numbers, becoming a single-particle relaxation effectively, and this system also has the same equilibrium state, $\rho_{ss}$. Here, we have two timescales in this case too, \textit{i.e.}, $T_1$ and $T_2$ along the $z$-direction and perpendicular to it, respectively. Therefore, if $T_1 = T_2$, relaxation is isotropic and we do not see MpE when there are no interactions among the particles or no shared environment, as discussed in the earlier section. For anisotropic relaxation, \textit{i.e.}, $T_1 > T_2$, the system has a intrinsic fast relaxation mode along the perpendicular direction to the $z$-axis; the system shows MpE and $T_2$ works as the timescale for MpE, as seen in Fig. \ref{fig:5}(a, b); though, unlike the system with a shared environment, the components of distance, \textit{i.e.}, $D_z$ and $D_\perp$, do not show crossover trajectories. However, this is also a ``genuine quantum MpE", as Fig. \ref{fig:5}(b) shows crossover trajectories for $S(\rho(t) || \rho_{ss})$ \cite{Moroder_2024}. For $T_2 \ll T_1$, the trajectories are almost perpendicular to the $z$-axis and the speed of convergence increases with distance from the mentioned axis, except on the $z$-axis where trajectories follow the axis towards $\rho_{ss}$ with speed $T_1^{-1}$, see Fig. \ref{fig:5}(d). Hence, the initial states satisfying $r_\circ \cos \theta_\circ = M_\circ$ thermalize early, and we see MpE, as shown in Fig. \ref{fig:5}(c). States lying on the mentioned line have zero overlap with the slowest relaxation mode and therefore bypass the corresponding long-time relaxation channel. Consequently, a state located on this line may relax faster even when it starts at a larger distance from equilibrium than another state that is initially closer to equilibrium but has a finite overlap with the slowest mode, as shown in Fig. \ref{fig:5}(a-b). Note that with increasing anisotropy, the system becomes more prone to exhibit MpE.

The Bloch equations employed in this work are derived within a mean-field framework, where higher-order correlations between different constituents are neglected and the dynamics is described solely in terms of single-particle expectation values. Consequently, quantitative aspects of the relaxation dynamics, such as the precise relaxation rates and the extent of the Mpemba regime in parameter space, may be modified when correlations are taken into account. Nevertheless, the emergence of the MpE in the present model is primarily associated with the nonlinear structure of the dynamics and the existence of multiple relaxation pathways characterized by distinct timescales. These features are captured within the mean-field description and are sufficient to generate the anomalous relaxation behavior characteristic of the MpE. Therefore, one expects the mean-field approximation to reproduce the essential mechanism responsible for the effect.

Though this simple relaxation dynamics provides a general model for MpE, it also suffers from the same drawback as any theory using mean-field approximations. The mean-field theories are nowhere near exact; they neglect the finer details of the system, like entanglement among the particles, fluctuations in the system, etc. This model primarily employs a semi-classical perspective of MpE, which is not a limitation; rather, it is advantageous, since the approach can be straightforwardly generalized to the classical MpE. A systematic investigation of the MpE in genuinely quantum relaxation dynamics, where many-body correlations and quantum coherence are fully retained, remains an important direction for future research.

The present work is based on a mean-field treatment of a spin-$1/2$ system, where the dynamics can be conveniently described using Bloch-sphere variables. Although higher-spin systems require a more general parametrization, the mean-field framework employed here can be extended in a similar fashion to such systems. As with the present model, this extension would provide a semi-classical description of the dynamics while neglecting effects beyond mean-field, such as correlations and fluctuations. Investigating the emergence of the MpE in higher-spin systems within this framework is a natural direction for future research, but lies beyond the scope of the present manuscript.

\section*{Conclusions}

This paper addresses the long-standing problem of identifying the origin of MpE in Markovian quantum systems. I demonstrate that MpE can emerge purely from relaxation dynamics, independent of the specific structure of the system or its environment. To this end, I consider a class of spin-$1/2$ systems in which the constituents are connected through all-to-all, equal-strength, energy-conserving interactions. Under a mean-field treatment, I show that the interaction strength does not affect the emergence of MpE; hence, this includes the class of systems consisting of noninteracting particles also. This insensitivity arises from the azimuthal symmetry of the relaxation dynamics. Furthermore, I show that when the constituents interact with a shared environment, collective dissipation generates a fast relaxation mode in the system. As the number of particles increases, this collective mode becomes more pronounced, leading to a stronger MpE. These results provide a physical mechanism through which multiple relaxation timescales emerge in spin-$1/2$ systems and suggest that collective behavior is the underlying origin of MpE in such cases. I also demonstrate that MpE arises naturally in systems with anisotropic relaxation, even in the absence of a shared environment. In these systems, the preferred direction of dissipation intrinsically gives rise to distinct relaxation timescales, making the occurrence of MpE increasingly likely as the degree of anisotropy grows. Being an open quantum dynamical model, this approach can naturally be extended to classical MpEs, thus providing a unified generic picture of MpE.

\textit{\textbf{Note} -} During the preparation of this manuscript, I became aware of a recent experiment by Chatterjee \textit{et al.} demonstrating the MpE in a Markovian system governed by dipolar relaxation dynamics \cite{Chatterjee_2025}.

\section*{Acknowledgments} 

I acknowledge the Anusandhan National Research Foundation - International Travel Support award (File No.: ITS/2025/003131) for supporting my travel to attend the ``School on Quantum Dynamics of Matter, Light and Information" (ICTP, Italy). I also thank the organizers of the mentioned school, where I was introduced to the Mpemba Effect (MpE) through the lectures of Pasquale Calabrese, during which Werner Krauth expressed his desire to know a general theory for MpE through relaxation dynamics; I would like to thank both of them. I want to thank Sarfraj Fency, Shubhamay Panja, Debanu Bandyopadhyay, and Rangeet Bhattacharyya for their helpful suggestions. I thank the anonymous reviewers for the helpful comments.

\bibliographystyle{apsrev4-2}
\bibliography{ref}

\appendix

\section{} \label{A:1}

In this section, I provide the correspondence between MpE in relative entropy ($S$) and Euclidean distance ($D$) for a spin-$1/2$ particle. For such a system, the density matrix can be represented using the magnetization as $\rho = \frac{1}{2} \left( \mathbb{I} + {\bm M } \cdot {\bm \sigma}\right)$ with $r = |\bm M| \in [0, 1]$ and equilibrium density matrix $\rho_{ss} = \frac{1}{2} \left( \mathbb{I} + M_\circ \sigma^z\right)$ with $M_\circ \in [-1,1]$. Therefore, we have the following,
\begin{eqnarray}
D \left( \rho, \rho_{ss} \right) &=& \sqrt{r^2 + M_\circ^2 - 2 M_\circ M_z} \\
S \left( \rho || \rho_{ss} \right) &=& \frac{1+r}{2} \ln \left( \frac{1+r}{2} \right) + \frac{1-r}{2} \ln \left( \frac{1-r}{2} \right) - \left( A+B \right) - \left( A-B \right) M_z
\end{eqnarray}
where, $A = \frac{1}{2} \ln \left( \frac{1+M_\circ}{2} \right)$ and $B = \frac{1}{2} \ln \left( \frac{1-M_\circ}{2} \right)$. This implies, $\frac{\partial D}{\partial r} = \frac{r}{D}$, $\frac{\partial D}{\partial M_z} = -\frac{M_\circ}{D}$, $\frac{\partial S}{\partial r} = \frac{1}{2}\ln \left( \frac{1+r}{1-r} \right)$, and $\frac{\partial S}{\partial M_z} = \frac{1}{2}\ln \left( \frac{1-M_\circ}{1+M_\circ} \right)$. Let's define a new variable $\nu$ as,
\begin{equation}
\nu = \begin{cases}
-M_z, & M_\circ \geq 0, \\
M_z, & M_\circ < 0.
\end{cases}
\end{equation}
Therefore, $S$ and $D$ are monotonically increasing functions of two independent variables $r$ and $\nu$, defined as,
\begin{eqnarray}
D \left( r, \nu \right) &=& \sqrt{r^2 + M_\circ^2 + 2 M_\circ \nu} \\
S \left( r, \nu \right) &=& \frac{1+r}{2} \ln \left( \frac{1+r}{2} \right) + \frac{1-r}{2} \ln \left( \frac{1-r}{2} \right) - \left( A+B \right) + \left( A-B \right) \nu
\end{eqnarray}
Hence, $S$ and $D$ preserve the order, \textit{i.e.} $S\left( r_1, \nu_1 \right) < S\left( r_2, \nu_2 \right) \Leftrightarrow D\left( r_1, \nu_1 \right) < D\left( r_2, \nu_2 \right)$. Therefore, MpE in $S$ necessarily and sufficiently imply MpE in $D$, for a given initial set of initial conditions for spin-$1/2$ systems. Consequently, the thermalization-time plots obtained using the Euclidean distance would be qualitatively similar to those shown in Figs. \ref{fig:3} and \ref{fig:5}, although they would not be exactly identical because D and S have different functional forms. Nevertheless, for spin-$1/2$ systems, the crossover of trajectories in D occurs at the same time as the crossover of the corresponding trajectories in S, and vice versa. Therefore, the ordering of thermalization times and the associated MpE conclusions remain unchanged, even though the numerical values of the thermalization times may differ.

\end{document}